\documentclass{mem}
\usepackage{natbib}\usepackage{txfonts}\usepackage{balance}
\usepackage{graphicx}
\usepackage[a4paper,breaklinks,dvipdfm]{hyperref}
\idline{75}{282}
\begin{document}
\def\teff{$T\rm_{eff }$}
\def\kms{$\mathrm {km s}^{-1}$}

\title{
Collisions versus stellar winds in the runaway merger scenario: place your bets
}

   \subtitle{}

\author{
Michela Mapelli 
          }

\institute{
Istituto Nazionale di Astrofisica --
Osservatorio Astronomico di Padova, Vicolo dell'Osservatorio 5,
I-35142 Padova, Italy, \email{michela.mapelli@oapd.inaf.it}
}

\authorrunning{Michela Mapelli}

\titlerunning{The runaway merger scenario revisited}

\abstract{
The runaway merger scenario is one of the most promising mechanisms to explain the formation of intermediate-mass black holes (IMBHs) in young dense star clusters (SCs). On the other hand, the massive stars that participate in the runaway merger lose mass by stellar winds. This effect is tremendously important, especially at high metallicity. We discuss N-body simulations of massive ($\sim{}6\times{}10^4$ M$_\odot$) SCs, in which we added new recipes for stellar winds and supernova explosion at different metallicity. At solar metallicity, the mass of the final merger product spans from few solar masses up to $\sim{}30$ M$_\odot$. At low metallicity ($0.01-0.1$ Z$_\odot$) the maximum remnant mass is $\sim{}250$ M$_\odot$, in the range of IMBHs. A large fraction ($\sim{}0.6$) of the massive remnants are not ejected from the parent SC and acquire stellar or black hole companions. Finally, I discuss the importance of this result for gravitational wave detection.
}
\maketitle{}

\section{Introduction}
GW150914 is the first direct detection of gravitational waves (GWs) ever \citep{abbott2016}. It has been interpreted as the coalescence of two black holes (BHs) with mass $m_1=36^{+5}_{-4}$ M$_\odot{}$ and $m_2=29^{+4}_{-4}$ M$_\odot$. This discovery has a plethora of implications for physics and astrophysics. Here, we will focus only on its implications for our understanding of stellar-mass BHs. GW150914 tells us that BH-BH binaries exist, can merge in a Hubble time, and they can be composed of massive stellar BHs, i.e. stellar BHs with mass $\ge{}25$ M$_\odot{}$ \citep{mapelli2009,mapelli2010,belczynski2010}. Hence, it is particularly important to study what processes can lead to the formation of massive stellar BHs or even intermediate-mass BHs (IMBHs, $\gtrsim{}100$ M$_\odot$). \cite{mapelli2009} proposed that metallicity is the essential key to understand the formation of massive stellar BHs: massive stars lose less mass by stellar winds if their are relatively metal poor ($Z\le{}0.5$ Z$_\odot$). Thus, their final mass (i.e. the mass before the core collapse) is higher than that of metal-rich stars with the same initial mass. According to several supernova (SN) models, if the final mass of a star is $\gtrsim{}30-40$ M$_\odot$, the SN can fail, and the star collapses to a BH directly. In case of a direct collapse, the mass of the BH can be significantly higher than in case of an explosion, because even the star envelope collapses. BHs with mass up to $\sim{}130$ M$_\odot$ can form via this mechanism (see \citealt{spera2015}, and references therein).

Stellar dynamics is another important key to understand the mass of BHs. Dynamics can affect the mass of BHs in several ways, triggering the merger of BHs with stars and with other BHs at different epochs (see e.g. \citealt{giersz2015}). According to the runaway collision scenario \citep{colgate1967,portegies2002,gurkan2004}, mass segregation is sufficiently fast in young star clusters (SCs), that massive stars can segregate to the centre before the first SNae explode. In the core, the most massive stars collide with each other, leading to the formation of a super-massive star ($>>100$ M$_\odot$), which might collapse into an IMBH. However, the final mass of the collision product strongly depends on mass loss by stellar winds. \cite{mapelli2016} studies the effect of stellar winds on the evolution of the collision product in the runaway scenario. This proceeding summarizes the main results of \cite{mapelli2016}.

\begin{figure}[t!]
\resizebox{\hsize}{!}{\includegraphics[width=4cm]{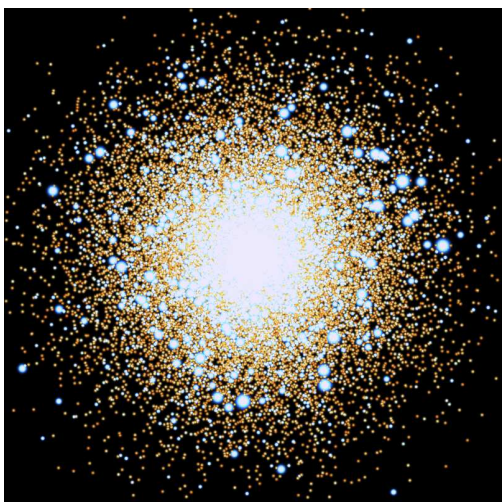}}
\caption{\footnotesize
Initial conditions of a simulated SC at $Z=1$ Z$_\odot$. 
The box length is 12 pc.
}
\label{fig:fig1}
\end{figure}

\begin{figure}[t!]
\resizebox{\hsize}{!}{\includegraphics[clip=true]{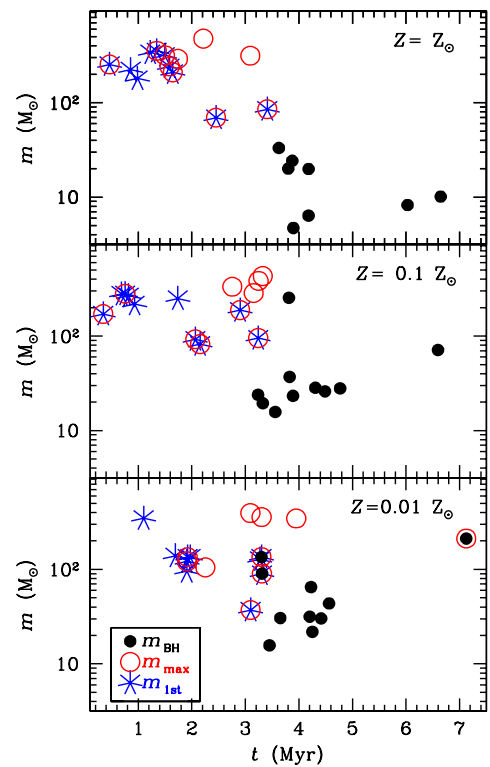}}
\caption{\footnotesize
Mass of the PCP as a function of time $t$. From top to bottom: $Z=1$, 0.1, 0.01 Z$_\odot$. Asterisks: mass of the PCP after the first collision. Open circles: maximum mass of the PCP.  Filled circles: mass of the BH born from the PCP.
}
\label{fig:fig2}
\end{figure}

\section{Methods}
We ran a set of 30 N-body realizations of young SCs with $N=10^5$ using the {\sc starlab} software environment, updated with the recipes for stellar evolution, stellar winds, and SN explosion presented in \cite{mapelli2013}. Ten of the simulations have metallicity $Z=0.01$, $0.1$ and $1$ Z$_\odot$, respectively. The SCs follow a King profile, with dimensionless central potential $W_0=9$, core radius $r_{\rm c}=0.05$ pc, and half-mass radius $r_{\rm hm}=1$ pc. Stellar masses follow the Kroupa initial mass function. No primordial binaries are included in the initial conditions. Thus, binaries form dynamically from gravitational encounters of three single bodies. We ran each SC in isolation for 17 Myr. Fig.~\ref{fig:fig1} shows the initial conditions of one of the simulations.

\begin{table}
\begin{center}
\caption{\label{tab:tab1}
Properties of the stable PCP binaries at the end of the simulations ($t=17$ Myr).}
 \leavevmode
\begin{tabular}[!h]{cccccc}
\hline
$Z$   & $M_{\rm PCP}$ & $M_{\rm co}$  & $P_{\rm orb}$ & $e$ & Type \\
(Z$_\odot$)   & (${\rm M}_\odot$) & (${\rm M}_\odot$) & (yr) & &\\
\hline
 0.01   & 32  & 19   & 1.63   & 0.41 & BH-BH \\
 0.01   & 22  & 5    & 0.0685 & 0.34 & BH-BH \\
 0.01   & 16  & 1.36 & 0.0160 & 0.22 & BH-NS \\
 0.01   & 212 & 47   & 4.50   & 0.35 & BH-BH \\
 0.01   & 135 & 64   & 1.37   & 0.92 & BH-BH \\
 0.1    & 19  & 38   & 3.67   & 0.72 & BH-BH \\
 0.1    & 254 & 38   & 0.60   & 0.39 & BH-BH \\
 1.0    & 20  & 21   & 9.67   & 0.31 & BH-BH \\
 1.0    & 5   & 19   & 19.3   & 0.15 & BH-BH \\
\noalign{\vspace{0.1cm}}
\hline
\end{tabular}
\begin{flushleft}
  \footnotesize{Column 1: metallicity ($Z$); column 2: mass of the PCP ($M_{\rm PCP}$); column 3: mass of the companion ($M_{\rm co}$); column 4: orbital period  ($P_{\rm orb}$); column 5: eccentricity  ($e$); column 6: binary type.}
\end{flushleft}
\end{center}
\end{table}
\section{Results and discussion}
In all simulations, we trace the history of collisions between stars. In particular, we name `principal collision product' (hereafter PCP) the product of the first collision occurring in each SC. We follow the evolution of the PCP through its subsequent mergers. Our aim is to find out what is the maximum mass the PCP can reach during the runaway collisions, and what is the mass of the dark remnant which forms from the collapse of the PCP. Each PCP undergoes from one up to six collisions, independent of metallicity. All collisions are triggered by three-body encounters with binaries. Due to the recoil velocity received during three-body encounters, $\sim{}40$ per cent of the PCPs are ejected from their parent SC.
\begin{figure}[]
\resizebox{\hsize}{!}{\includegraphics[]{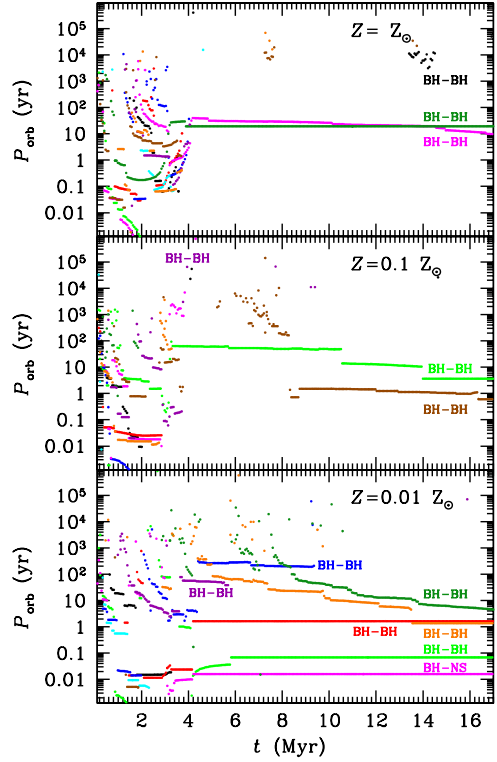}}
\caption{
\footnotesize
Orbital period of the binary systems whose member is a PCP as a function of time. From top to bottom: $Z=$ Z$_\odot$, $Z=0.1$ Z$_\odot$ and $Z=0.01$ Z$_\odot$. Each line is a single PCP from a different SC simulation. 
}
\label{fig:fig3}
\end{figure}

Fig.~\ref{fig:fig2} shows the mass of the PCPs as a function of time. The maximum mass a PCP star can reach in our simulations is $\sim{}500$ M$_\odot$, nearly independent of the metallicity. However,  a 500 M$_\odot$ star loses mass very fast by stellar winds at $Z\sim{}$ Z$_\odot$. Thus, the mass of the BH born from the PCP at $Z=$ Z$_\odot$ is always $<40$ M$_\odot$. In contrast, mass loss is less effective at lower metallicity ($Z=0.01,$ 0.1 Z$_\odot$). This explains why four of the PCP BHs  at $Z=0.01$ and 0.1 Z$_\odot$ have mass $m_{\rm BH}>90$ M$_\odot$, i.e. they can be considered IMBHs. The maximum mass of IMBHs born from runaway collisions in our simulations is $m_{\rm BH}\sim{}250$ M$_\odot$. However, this maximum mass strongly depends on the recipes of stellar winds that we adopted and should be regarded as quite optimistic.

We then study whether the PCPs acquire stellar or BH companions. Fig.~\ref{fig:fig3} shows the orbital period of the PCP binaries (i.e. those binaries including a PCP) as a function of time. At early epochs ($t<4$ Myr) PCP binaries  form dynamically, harden by three-body encounters, and occasionally break by merger or SN explosion. At later times ($t>4$ Myr) only few hard PCP binaries survive. Interestingly, all hard and stable PCP binaries are double compact object binaries. Four stable BH-BH binaries plus one stable neutron star (NS)-BH binary form at $Z=0.01$ Z$_\odot$, while two stable BH-BH binaries form at both $Z=0.1$ Z$_\odot$ and $Z=Z_\odot$. The final orbits of such binaries range from few days to few years (Table~\ref{tab:tab1}). Interestingly, some of the orbital periods are still decreasing at the end of the simulations (17 Myr), because of hardening and dynamical exchanges. Eccentricities range from $\sim{}0.15$ up to $\sim{}0.9$ (Table~\ref{tab:tab1}). Large eccentricities indicate recent exchanges. Three BH-BH binaries at low $Z$ 
contain IMBHs, with a mass ranging from $\sim{}130$ to $\sim{}250$ M$_\odot$ (Table~\ref{tab:tab1}). Finally, two binaries 
have masses similar to the ones of the progenitors of GW150914 (see the first and the sixth line of Table~\ref{tab:tab1}). 

A simple estimate of the coalescence timescale due to orbital decay by GW emission is given by
\begin{equation}
t_{\rm GW}=\frac{5}{256}\,{}\frac{c^5\,{}a^4\,{}(1-e^2)^{7/2}}{G^3\,{}m_1\,{}m_2\,{}(m_1+m_2)}, 
\end{equation}
where $c$ is the speed of light, $G$ the gravitational constant, $a$ the semi-major axis of the binary, $e$ the eccentricity, and $m_1$ ($m_2$) the mass of the primary (secondary) member of the binary \citep{peters1964}. If we estimate $t_{\rm GW}$ for the binaries reported in Table~\ref{tab:tab1}, we find that $t_{\rm GW}>500$ Gyr for all systems, i.e. none of the PCP binaries is expected to merge in a Hubble time. However, this is a clear lower limit to the merger rate, because if any of the binaries manages to remain in the SC for more than $17$ Myr (i.e. the duration of the simulation) its coalescence timescale is expected to decrease further by hardening and exchanges.

Finally, if we consider all double compact-object binaries in our simulations (not only those containing a PCP), we find one BH-BH system, with masses $m_1=17$ M$_\odot$, $m_2=16$ M$_\odot$, which is expected to merge in a Hubble time ($t_{\rm GW}\sim{}1.2$ Gyr). Using the same formalism as described in equation~3 of \cite{ziosi2014}, we derive an expected merger rate of $R\sim{}10^{-3}$ Gpc$^{-3}$ yr$^{-1}$, close to the lower limit of the merger rate estimated from LIGO's detection \citep{abbott2016}. Our estimated merger rate is a pessimistic value for several reasons: first, we simulate the SCs only for 17 Myr; second, we did not include  primordial binaries. We will account for these effects in forthcoming studies.

\section{Summary}
We revisited the runaway collision scenario in SCs by accounting for metallicity-dependent stellar winds and SN explosions. Massive stars with mass up to $\sim{}500$ M$_\odot$ form from the runaway merger, regardless of SC metallicity. At solar metallicity, the final mass of the BH born from the collision product is always $<40$ M$_\odot$. At lower metallicity (0.01, 0.1 Z$_\odot{}$), BHs with mass up to $\sim{}250$ M$_\odot$ form in our simulations. However, the maximum BH mass is very sensitive to our assumptions about mass loss. All stable binaries including collision products are double compact-object binaries. This result is of foremost importance for the study of GW sources and deserves further study.



\begin{acknowledgements}
MM acknowledges financial support from MIUR through grant FIRB 2012 RBFR12PM1F, from INAF through grant PRIN-2014-14, and from the MERAC Foundation.
\end{acknowledgements}

\bibliographystyle{aa}

\end{document}